\documentclass[aps,prd,twocolumn,showkeys,nofootinbib]{revtex4}
\pdfoutput=1
\usepackage{epsfig}
\usepackage{amsmath}
\usepackage{color}
\usepackage{graphicx}
\newcommand{\beq}{\begin{equation}}
\newcommand{\eeq}{\end{equation}}
\newcommand{\bqa}{\begin{eqnarray}}
\newcommand{\eqa}{\end{eqnarray}}

\begin{document}

\title{Doubly charmed tetraquarks  in a diquark-antidiquark model }
\author{Xiaojun Yan, Bin Zhong and Ruilin Zhu~\footnote{Email:rlzhu@njnu.edu.cn}  }
\affiliation{
Department of Physics and Institute of Theoretical Physics,
Nanjing Normal University, Nanjing, Jiangsu 210023, China
 }

\begin{abstract}
We study the spectra of the doubly charmed tetraquark states  in a diquark-antidiquark model.
 The doubly charmed tetraquark states form an antitriplet and a sextet configurations according to flavor SU(3) symmetry. For the tetraquark state $[qq'][\bar c\bar c]$, we show the mass for both bound and excited states. The two-body  decays of tetraquark states $T^{cc}[0^+]$ and $T^{cc}[1^{--}]$ to charmed mesons have also been studied. In the end,the doubly charmed tetraquarks decays to a charmed baryon and a  light baryon have been studied in the SU(3) flavor symmetry.

\keywords{Exotic states, tetraquark, diquark }

\end{abstract}

\maketitle

\section{Introduction}

Since last decade, Belle, $BABAR$, BESIII, LHCb and other experimental data have indicated a considerable
number of exotic hadronic resonances with charm or beauty, the so-called $XYZ$ and $P_c$ particles including tetraquark states and pentaquark states~\cite{Choi:2003ue,Aubert:2004fc,Aubert:2005rm,Wang:2007ea,Yuan:2007sj,Aubert:2007zz,Choi:2007wga,Pakhlova:2008vn,Mizuk:2008me,Aaltonen:2009tz,
Uehara:2009tx,Ablikim:2013mio,Ablikim:2013wzq,Ablikim:2013xfr,Ablikim:2013emm,D0:2016mwd,Aaij:2016iza}. All of these exotic states have many unexpected properties, such as masses, decay widths and cross-sections, which are hard to be explained in the conventional quark model. The nature of these states is one of the most interesting subjects in hadron physics.

The study of heavy flavor tetraquark states consisting of two quarks and two antiquarks has a long history. Hidden charm and bottom tetraquark states have been investigated in the past decades~\cite{Ader:1981db,Ballot:1983iv,Heller:1985cb,Badalian:1985es,Chen:2004dy,Cheng:2003kg,Maiani:2004vq,Ali:2009pi,Ali:2010pq,Chen:2010ze,Ali:2011ug,
Karliner:2013dqa,Ali:2014dva,Liu:2014dla,Karliner:2015ina,Maiani:2014aja,Ma:2014zva,
Wang:2016mmg,Lebed:2016yvr,Wang:2016tsi,He:2016xvd}. The recent reviews of  the exotic states can be found in Refs.~\cite{Chen:2016qju,Esposito:2016noz,Guo:2017jvc}. Currently, the exact inner  physical picture of these exotic states have not been reached an agreement in both the theoretical and experimental communities.  The possible explanations of the  $XYZ$ resonances can  be classified into hadronic molecules, compact tetraquarks, diquark-antidiquark states, conventional quarkonium, triangle anomaly, and kinematics effects.

In Ref.~\cite{Esposito:2016noz}, the multiquark resonances are reviewed and a coherent description of the so called X and Z resonances is presented. The first suggestion to use light diquarks to explain the exotic states is from Jaffe and Wilczek~\cite{Jaffe:2003sg}. Maiani et al.  then introduced heavy-light diquarks to explain the charmonium-like states in  Ref.~\cite{Maiani:2004vq}. The evidence that in a tetraquark system the two quarks arrange their color in a diquark before interacting with the antiquarks has been found on the lattice~\cite{Cardoso:2011fq}.

 Based on the diquark-antidiquark model,  tetraquark states $[cq'][\bar c \bar q]$ can form an octet representation and a singlet representation in flavor SU(3) symmetry which may give an interpretation to such $XYZ$ states~\cite{Zhu:2016arf}. A tetraquark state can also be represented with four quark configuration $[qq'][\bar c\bar c]$(two charm antiquarks $\bar c$'s and two light quarks with up, down and strange quarks $u$, $d$ and $s$), which is called as doubly charmed (C=2) tetraquark $T^{cc}$. From a theoretical point of view, $T^{cc}$ may be a  more interesting hadronic state because
the two heavy charm quark are more likely to form a lower energy diquark and two light quarks will rotate round the charmed diquark. Besides, its flavor quantum numbers can give the possibility to clearly distinguish these tetraquarks from conventional quarkonia.

The doubly bottomed tetraquark states  with four quark configuration $[bb][\bar q\bar q']$(two bottom quarks and two light antiquarks) have been studied in Refs.~\cite{Karliner:2017qjm,Eichten:2017ffp,Francis:2016hui,Bicudo:2016ooe,Mehen:2017nrh}. In Ref.~\cite{Karliner:2017qjm}, a doubly bottomed tetraquark
with spin-parity $J^P=1^+$ and mass $m=10389$MeV was predicted, which is around 215 MeV below $B\bar{B}^* $ threshold. In Ref.~\cite{Eichten:2017ffp}, a doubly bottomed tetraquark
around 135 MeV below $B\bar{B}^* $ threshold was predicted. The Lattice calculations were performed in Refs.~\cite{Francis:2016hui,Bicudo:2016ooe}, where the doubly bottomed tetraquark below $B\bar{B}^* $ threshold were also predicted. In Ref.~\cite{Mehen:2017nrh}, the doubly bottomed and charmed tetraquarks
were discussed in the heavy quark symmetry.

In this paper, we discuss the possibility of the doubly charmed tetraquark states, and attempt to investigate  the mass spectrum and decay widths of doubly charmed tetraquarks in a diquark-antidiquark model. The doubly charmed tetraquark states can form an antitriplet  configuration and a sextet configuration according to flavor SU(3) symmetry. We hope  the  doubly charmed tetraquark  states $T^{cc}$ can be discovered at BES III, Belle II, LHCb and other experiments with their large data samples of high luminosity.

The paper is organized as the following. In Sec.~\ref{II}, we classify the doubly charmed tetraquark states into anti-triplet and sextet representations in  flavor SU(3) symmetry.
 In Sec.~\ref{III}, the spectra  of the doubly charmed tetraquark states  are calculated from  a diquark-antidiquark model. In Sec.~\ref{IV}, we study
 the two-body charmed mesons decays of  doubly charmed tetraquarks. We summarize and conclude in the end.

\section{Doubly  charmed  tetraquarks in flavor SU(3) symmetry\label{II}}

Considering the flavor $SU(3)$ symmetry, the doubly charmed tetraquarks $T^{cc}\sim [qq'][\bar{c}\bar{c}]$ with two light quarks, $q$ and $q'$, can be conventionally classified into two groups.
The three light quarks, $(u,\; d,\; s)$ form a triplet ${\bf 3}$ representation and the charm quark $c$ is a singlet~\cite{Zeppenfeld:1980ex,Chau:1990ay,Gronau:1994rj}. The doubly charmed tetraquarks $T^{cc}\sim [qq'][\bar{c}\bar{c}]$ can have the following irreducible representations
\begin{eqnarray}
{\bf 3} \otimes  {\bf 3} = {\bf \bar 3}\oplus {\bf 6}\;.
\end{eqnarray}

 We label ${\bf \bar 3}$ representation by $T^{cc}_{[i,j]}$. Here the flavor components are antisymmetric under the exchange of $i$ and $j$, and thus $T^{cc}_{[i,j]}$ is traceless as $T^{cc}_{[i,i]} = 0$. The components can be given explicitly as
\begin{eqnarray}
 &&T^{cc}_{[1,2]}= \frac{1}{\sqrt 2} T^{cc}_{ud}({\bf \bar 3}), \;\;\;\;\; T^{cc}_{[1,3]}= \frac{1}{\sqrt 2} T^{cc}_{us}({\bf \bar 3}),  \nonumber\\
 && T^{cc}_{[2,3]}= \frac{1}{\sqrt 2} T^{cc}_{ds}({\bf \bar 3}).
\end{eqnarray}

We label ${\bf 6}$ representation by $T^{cc}_{\{i,j\}}$. Here the flavor components are symmetric under the exchange of $i$ and $j$. In this case,  the components are given by
\begin{eqnarray}
 T^{cc}_{\{1,2\}}= \frac{1}{\sqrt 2} T^{cc}_{ud}({\bf 6}), &&\;\;\;\;\; T^{cc}_{\{1,3\}}= \frac{1}{\sqrt 2} T^{cc}_{us}({\bf 6}),  \nonumber\\
  T^{cc}_{\{2,3\}}= \frac{1}{\sqrt 2} T^{cc}_{ds}({\bf 6}),&&\;\;\;\;\; T^{cc}_{\{1,1\}}=  T^{cc}_{uu}({\bf 6}),\nonumber\\
 T^{cc}_{\{2,2\}}=  T^{cc}_{dd}({\bf 6}),&&\;\;\;\;\; T^{cc}_{\{3,3\}}=  T^{cc}_{ss}({\bf 6}),
\end{eqnarray}

To summarize, the flavor components of tetraquarks in flavor SU(3) symmetry can be explicitly obtained as below
\begin{eqnarray}
 T^{cc}_{ud}({\bf \bar 3})= \frac{1}{\sqrt 2} (ud-d u)\bar c\bar c, &&\;\;\;  T^{cc}_{us}({\bf \bar 3})= \frac{1}{\sqrt 2} (us-s u)\bar c\bar c, \nonumber\\
  T^{cc}_{ds}({\bf \bar 3})= \frac{1}{\sqrt 2} (ds-s d)\bar c\bar c, &&\;\;\;  T^{cc}_{ud}({\bf 6})= \frac{1}{\sqrt 2} (ud+d u)\bar c\bar c, \nonumber\\
   T^{cc}_{us}({\bf 6})= \frac{1}{\sqrt 2} (us+s u)\bar c\bar c, &&\;\;\;  T^{cc}_{ds}({\bf 6})= \frac{1}{\sqrt 2} (ds+sd)\bar c\bar c, \nonumber\\
    T^{cc}_{uu}({\bf 6})= uu\bar c\bar c, &&\;\;\;    T^{cc}_{dd}({\bf 6})= dd\bar c\bar c, \nonumber\\
     T^{cc}_{ss}({\bf 6})= ss\bar c\bar c.&&
\end{eqnarray}

The orbitally excited tetraquark states also form anti-triplet and sextet representations. Considering the first orbital excitation with $L=1$,
the orbitally excited tetraquark states will have the spin-parity $1^-$.
For the neutral tetraquarks, $T^{cc}_{uu}$ can have the definite charge-parity,
thus their $J^{PC}$ quantum numbers can be $1^{--}$ or $1^{-+}$.

\section{Doubly  charmed  tetraquarks  spectra\label{III}}
Because the heavy charm quarks interact with each other at the momentum scale  $m_c v$ with the heavy charm quark relative velocity $v$,  and this scale  $m_c v$ is  a large quantity compared with the typical hadron scale, the two heavy charm quarks will have a small distance between each other and are easily to form an attractive diquark.
Considering a diquark-antidiquark picture $\delta \bar \delta'$ with $\delta=[q q']$ and $ \bar \delta'=[\bar c \bar c]$, the effective Hamiltonian includes three kinds of interactions: spin-spin interactions of quarks in the diquark, antidiquark and between them;  spin-orbital interactions;  orbit-orbital interactions. The effective Hamiltonian then can be written as~\cite{Maiani:2004vq}:
\begin{eqnarray}
 H&=&m_{\delta}+m_{\delta^\prime}+H^{\delta}_{SS} + H^{\bar{\delta^\prime}}_{SS}+H^{\delta\bar{\delta^\prime}}_{SS} + H_{SL}+H_{LL},\nonumber\\
 \label{eq:definition-hamiltonian}
\end{eqnarray}
where $m_\delta$ and $m_{\delta^\prime}$ are the constituent masses of the diquark $[qq^\prime]$ and the antidiquark $[\bar{c}\bar{c}]$, respectively. $H^\delta_{SS}$ and $H^{\bar{\delta^\prime}}_{SS}$ denote the spin-spin interaction inside the diquark and antidiquark, respectively. $H^{\delta\bar{\delta^\prime}}_{SS}$ denotes the spin-spin interaction of quarks between diquark and antidiquark. $H_{SL}$ and $H_{LL}$ represent the spin-orbital and purely orbital interactions.

The explicit form of each Hamiltonian is written as
\begin{eqnarray}
 H^\delta_{SS}&=&2(\kappa_{q q'})_{\bar{3}}(\mathbf{S}_q\cdot \mathbf{S}_{q'}),\nonumber\\
 H^{\bar{\delta^\prime}}_{SS}&=&2(\kappa_{{\bar c \bar c}})_{\bar{3}}(\mathbf{S}_{\bar{c}}\cdot \mathbf{S}_{\bar{c}}), \nonumber\\
 H^{\delta\bar{\delta^\prime}}_{SS} &=&4\kappa_{q'\bar{c}}(\mathbf{S}_{q'}\cdot \mathbf{S}_{\bar{c}}) +4\kappa_{q \bar{c}}(\mathbf{S}_{q}\cdot \mathbf{S}_{\bar{c}}),
\nonumber\\
 H_{SL}&=&2 A_\delta (\mathbf{S}_\delta \cdot \mathbf{L}) +2 A_{\bar{\delta^\prime}}(\mathbf{S}_{\bar{\delta^\prime}}\cdot \mathbf{L}),\nonumber\\
 H_{LL}&=&B_{\delta\bar{\delta^\prime}} \frac{L(L+1)}{2}\ .
\label{eq:definition-hamiltonian2}
\end{eqnarray}
where $\mathbf{S}_{q(q')}$ and $\mathbf{S}_{c(\bar c)}$ are the spin operators of light and heavy quarks, respectively. $\mathbf{S}_{\delta}$ and $\mathbf{S}_{\bar{\delta^\prime}}$ denote the spin operators  of the diquark and antidiquark, respectively. $\mathbf{L}$ is the orbital angular momentum operator. The other parameters are all coefficients. $A_{\delta(\bar{\delta^\prime})}$ and $B_{\delta\bar{\delta^\prime}}$ are the spin-orbit and orbit-orbit couplings, respectively.
$(\kappa_{q q'})_{\bar{3}}$ and $(\kappa_{\bar c \bar c})_{\bar{3}}$ are the spin-spin couplings for diquark in color antitriplet ${\bf\bar 3}$. $\kappa_{q' \bar c}$ and $\kappa_{q \bar c}$ are the spin-spin couplings for a quark-antiquark pair.

The orbital angular momentum of ground states of tetraquark is zero.  In this case, there are two possible tetraquark configurations  with the spin-parity $J^P=0^+$ in the spin space,
\begin{eqnarray}
|0_\delta,0_{\bar{\delta^\prime}},0_J\rangle&=&\frac{1}{2}
\big[(\uparrow)_q(\downarrow)_{q^\prime}-(\downarrow)_q(\uparrow)_{q^\prime} \big](\uparrow)_{\bar{c}}(\downarrow)_{\bar{c}}
\nonumber\\
&&-\frac{1}{2}
\big[(\uparrow)_q(\downarrow)_{q^\prime}-(\downarrow)_q(\uparrow)_{q^\prime} \big](\downarrow)_{\bar{c}}(\uparrow)_{\bar{c}},\nonumber\\
|1_\delta,1_{\bar{\delta^\prime}},0_J\rangle&=&\frac{1}{\sqrt{3}}
\big\{(\uparrow)_q(\uparrow)_{q^\prime}(\downarrow)_{\bar{c}}(\downarrow)_{\bar{c}}
+(\downarrow)_q(\downarrow)_{q^\prime}(\uparrow)_{\bar{c}}(\uparrow)_{\bar{c}} \nonumber\\&& -\frac{1}{2}
\big[(\uparrow)_q(\downarrow)_{q^\prime}+(\downarrow)_q(\uparrow)_{q^\prime} \big](\uparrow)_{\bar{c}}(\downarrow)_{\bar{c}}
\nonumber\\&& -\frac{1}{2}
\big[(\uparrow)_q(\downarrow)_{q^\prime}+(\downarrow)_q(\uparrow)_{q^\prime} \big](\downarrow)_{\bar{c}}(\uparrow)_{\bar{c}}\big\}.
 \label{eq:definition-states0+}
\end{eqnarray}
where $|S_\delta,S_{\bar{\delta^\prime}},S_J\rangle $ denotes the doubly charmed tetraquark; the $S_\delta$ and $S_{\bar{\delta^\prime}}$ represent  the spin  of diquark $[qq^\prime]$ and antidiquark $[\bar{c}\bar{c}]$, respectively; the $S_J$ represents the total angular momentum of the tetraquark.

The corresponding splitting mass matrix for the $J^P=0^+$ tetraquarks is
\begin{eqnarray}
\Delta M (0^+)=\left(
\begin{array}{cc}
 -\frac{3}{2} ((\kappa_{q{q^\prime}})_{\bar{3}}+(\kappa_{{\bar c \bar c}})_{\bar{3}}) & 0\\
 0& h_1
\end{array} \right),
\end{eqnarray}
where
\begin{eqnarray}
h_1&= &\frac{1}{2} ((\kappa_{q{q^\prime}})_{\bar{3}}+(\kappa_{{\bar c \bar c}})_{\bar{3}}-4 \kappa_{{q \bar{c}}}-4\kappa_{{q'\bar{c}}}).\nonumber\\
\end{eqnarray}
The mass matrix is given as
\begin{eqnarray}
 M(J^P)= m_{\delta}+m_{\delta'}+\Delta M(J^P).
\end{eqnarray}
 The above mass matrix is naturally diagonalized, and one can easily obtain two different eigenvalues.

The corresponding tetraquark configuration for $J^P=2^+$ is
\begin{eqnarray}
|1_\delta,1_{\bar{\delta^\prime}},2_J\rangle&=&
(\uparrow)_q(\uparrow)_{q^\prime}(\uparrow)_{\bar{c}}(\uparrow)_{\bar{c}},
\label{eq:definition-states2+}
\end{eqnarray}
with the mass
\begin{eqnarray}
 M(2^+)&=&m_\delta+m_{\delta^\prime}+\frac{1}{2}\left( (\kappa_{q{q^\prime}})_{\bar{3}}+(\kappa_{{\bar c \bar c}})_{\bar{3}}\right)\nonumber\\&&+\frac{1}{2}\left(2\kappa_{{q\bar{c}}}+2\kappa_{{q'\bar{c}}}\right).
\end{eqnarray}

 The possible configurations  for the  tetraquark with $J^P=1^+$ are
\begin{eqnarray}
|0_\delta,1_{\bar{\delta^\prime}},1_J\rangle&=&\frac{1}{\sqrt{2}}
\big[(\uparrow)_q(\downarrow)_{q^\prime}-(\downarrow)_q(\uparrow)_{q^\prime} \big](\uparrow)_{\bar{c}}(\uparrow)_{\bar{c}}
 ,\nonumber \\
 |1_\delta,0_{\bar{\delta^\prime}},1_J\rangle&=&\frac{1}{\sqrt{2}}
(\uparrow)_q(\uparrow)_{q^\prime}\big[(\uparrow)_{\bar{c}}(\downarrow)_{\bar{c}}-(\downarrow)_{\bar{c}}
(\uparrow)_{\bar{c}}\big] \nonumber\\
|1_\delta,1_{\bar{\delta^\prime}},1_J\rangle&=&\frac{1}{2}
\big\{(\uparrow)_q(\uparrow)_{q^\prime}\big[(\uparrow)_{\bar{c}}(\downarrow)_{\bar{c}}+(\downarrow)_{\bar{c}}
(\uparrow)_{\bar{c}}\big]\nonumber\\&&-\big[(\uparrow)_q(\downarrow)_{q^\prime}+(\downarrow)_q(\uparrow)_{q^\prime} \big](\uparrow)_{\bar{c}}(\uparrow)_{\bar{c}}\big\}.
 \label{eq:definition-states1+}
\end{eqnarray}
The tetraquarks with the quark content $[qq'][\bar{c}\bar{c}]$ do not have any definite charge parity if $q\neq u$ and $q'\neq u$. Thus the above three $1^+$ states can mix with each other.

The mass splitting matrix $\Delta M$ for $J^P=1^+$ can be obtained with the base vectors defined in Eq. (\ref{eq:definition-states1+}).

\begin{widetext}
\begin{eqnarray}
\Delta M=\left(
\begin{array}{ccc}
 \frac{1}{2} ((\kappa_{{\bar c \bar c}})_{\bar{3}}-3 (\kappa_{q{q'}})_{\bar{3}}) & 0 & \sqrt{2}(\kappa_{{q'\bar{c}}}-\kappa_{{q\bar{c}}}) \\
 0 & \frac{1}{2} ((\kappa_{q{q^\prime}})_{\bar{3}}-3 (\kappa_{{\bar c \bar c}})_{\bar{3}}) & 0 \\
  \sqrt{2}(\kappa_{{q'\bar{c}}}-\kappa_{{q\bar{c}}}) & 0 & \frac{1}{2}
   ((\kappa_{q{q'}})_{\bar{3}}+(\kappa_{{\bar c \bar c}})_{\bar{3}}-2\kappa_{{q' \bar{c}}}-2\kappa_{{q \bar{c}}})
\end{array}
\right),\nonumber\\
\end{eqnarray}
\end{widetext}

In flavor SU(3) symmetry, all tetraquark states without orbital angular momenta should have the identical mass.   Since the strange quark is different from the up and down quarks, the reasonable tetraquark masses could be obtained.  In the calculation of the doubly charmed tetraquark spectra, we will distinguish the strange quark from the up and down quarks.
For the light diquark, $m_{qq}$ is determined  to be 0.395GeV by the $f_0(500)$ particle~\cite{Maiani:2004vq}, $m_{sq}$ is determined  to be 0.590GeV by the $a_0(980)$ particle~\cite{Maiani:2004vq}, and the strange diquark mass $m_{ss}=0.785${GeV} is estimated by the relation $m_{ss}-m_{sq}=m_{sq}-m_{qq}$~\cite{Zhu:2015bba,Wang:2016tsi}, where $q$ stands for $u$ or $d$.
The heavy antidiquark mass $m_{{\bar c\bar c}}$ is estimated by the relation $m_{{\bar c\bar c}}\simeq 2m_{c}$, since the heavy quark is highly static in the rest frame of the hadron.
In the paper,
we adopt the heavy quark mass as $ ~m_c=1.670\mathrm{GeV}$~\cite{Maiani:2004vq,Wang:2016tsi}.

Different diquark masses will obviously affect the tetraquark's mass.   In Refs.~\cite{Oettel:1998bk,Nicmorus:2008vb,Eichmann:2016yit}, the diquark masses effects are studied.  The scalar and axial-vector diquark masses are assumed to be equal in order to limit the number of parameters.
The diquark mass is chosen to be $m^{fg}_\delta=\xi(m_f+m_g)$ where $fg\in \{ qq, qs, ss\}$ is the flavor content of the diquark. Naturally, the diquark mass parameter $\xi$ is assumed to be $\xi\in[0,1]$~\cite{Oettel:1998bk}. In order to consider the effects from the diquark mass, we denote the running diquark mass $m^{fg}_\delta$ and we have $m^{fg}_\delta=m_\delta+\Delta\delta$. Thus we have $\Delta\delta=\xi(m_f+m_g)-m_\delta$.

For the spin-spin couplings, the strange quark is also treated differently from the up and down quarks~\cite{Maiani:2004vq,Wang:2017vnc,Zhu:2015bba}. The couplings are chosen as: $(\kappa _{qq})_{\bar 3}=103$MeV, $(\kappa _{sq})_{\bar 3}=64$MeV, $(\kappa _{{\bar c\bar c}})_{\bar 3}=39$MeV, $(\kappa _{{q\bar{c}}})_{0}=70$MeV and $(\kappa_{{s\bar{c}}})_{0}=72$MeV where $q$ also stands for $u$ or $d$. The relation $\kappa_{ij}=\frac{1}{4}(\kappa_{ij})_{0}$ is satisfied  for the quark-antiquark state, which is extracted from one gluon exchange model. The spin-orbit
coupling $A_{\delta}$ is  estimated as 30MeV, and the orbit-orbit coupling
$B_{\delta \bar{\delta}'}$  is estimated as $278~\mbox{MeV}$~\cite{Maiani:2004vq,Zhu:2015bba}.

 The wave function of a tetraquark  consists of four parts: space-coordinate, color, flavor, and spin subspaces,
 \begin{eqnarray}
\Psi(q,q',\bar{c},\bar{c})&=&\psi(x_1,x_2,x_3,x_4)\otimes
\chi_c(c_1,c_2,c_3,c_4)\nonumber\\&&\otimes \chi_f(f_1,f_2,f_3,f_4)\otimes \chi_s(s_1,s_2,s_3,s_4)\,,\nonumber\\
\end{eqnarray}
where $\psi(x_i)$, $\chi_c(c_i)$, $\chi_f(f_i)$, and $\chi_s(s_i)$ denote the space, color, flavor, and spin wave functions, respectively. The sub-labels 1, 2, 3, 4 denote $q$, $q'$, $\bar{c}$, $\bar{c}$, respectively.
 The  diquark is attractive only in the triplet representation in color space, thus the color wave function is antisymmetric.  The  antidiquark $[\bar{c}\bar{c}]$ is symmetric in flavor space, thus the spin wave function of the antidiquark $[\bar{c}\bar{c}]$ has to be symmetric with $S_{\bar{\delta^\prime}}=1$ when we do not consider the orbital excitation in the inner diquark system.

First we focus on the tetraquarks with $L=0$,
and then the space wave function is symmetric.  If the spin wave function of the diquark system $[qq']$ is antisymmetric, i.e. $S_{\delta}=0$, the flavor function should be also antisymmetric. In this case, the charmed  tetraquarks  can be decomposed into the ${\bf\bar 3}$ representation, with  the spin-parity $J^P=1^+$. Inputting the parameter masses and using the fixed diquark mass with $\Delta\delta=0$, their masses are  determined  to be
\begin{align}\label{mass1}
 m(T^{cc}_{ud}({\bf \bar 3}))&= 3.60 {\rm GeV} , & J^P=1^+ ,
 \\
  m(T^{cc}_{us}({\bf \bar 3}))=  m(T^{cc}_{ds}({\bf \bar 3}))&= 3.85 {\rm GeV} , & J^P=1^+ .
\end{align}
 If using the running diquark mass with $\Delta\delta\neq 0$, the tetraquark mass  $m(T^{cc}_{qq'})$ will become to be  $m(T^{cc}_{qq'})+\Delta\delta$. In Eqs.~(\ref{mass1}-\ref{mass3}),
we have assumed $\Delta\delta= 0$.

If the spin wave function of the diquark system $[qq']$ is symmetric, i.e. $S_{\delta}=1$, the flavor function should be also symmetric. In this case, the charmed  tetraquarks  can be decomposed into the ${\bf 6}$ representation, with  the spin-parity $J^P=0^+,1^+,2^+$. Their masses are determined   to be
\begin{align}\label{mass2}
 m(T^{cc}_{ud}({\bf 6}))&= \left\{ \begin{array} {ll}
  3.94 {\rm GeV} , & J^P=0^+ , \\
3.97 {\rm GeV} , & J^P=1^+ ,
 \\ 4.04 {\rm GeV}, & J^P=2^+ ,\end{array} \right.
 \end{align}
 \begin{align}
  m(T^{cc}_{us}({\bf 6}))&= \left\{ \begin{array} {ll}
 4.11 {\rm GeV} , & J^P=0^+ , \\
4.15{\rm GeV} , & J^P=1^+ ,
 \\ 4.22 {\rm GeV}, & J^P=2^+ ,\end{array} \right.\\
  m(T^{cc}_{ss}({\bf 6}))&= \left\{ \begin{array} {ll}
  4.30{\rm GeV} , & J^P=0^+ , \\
4.34 {\rm GeV} , & J^P=1^+ ,
 \\ 4.41 {\rm GeV}, & J^P=2^+ .\end{array} \right.
\end{align}

Other tetraquarks' masses can be obtained by $  m(T^{cc}_{uu}({\bf 6}))=  m(T^{cc}_{dd}({\bf 6}))=  m(T^{cc}_{ud}({\bf 6}))$ and $m(T^{cc}_{ds}({\bf 6}))=  m(T^{cc}_{us}({\bf 6}))$.

From the above calculation, a doubly charmed tetraquark $T^{cc}_{ud}({\bf \bar 3})$ with spin-parity $J^P=1^+$ and mass $M=3.60$GeV is predicted in a diquark-antidiquark model, which is around 140MeV below the ${D\bar D}$ threshold and 270MeV below the ${D\bar D^*(D^*\bar D)}$ threshold. Other doubly charmed tetraquark  states are predicted above the ${D\bar D}$ threshold, which will have large decay widths compared with the tetraquark $T^{cc}_{ud}({\bf \bar 3})$  with spin-parity $J^P=1^+$ and mass $3.60$GeV.

For the orbitally excited  tetraquark states with $L=1$, here we just consider the tetraquarks with spin-parity $J^P=1^-$.
The masses of the tetraquarks with  the spin-parity $J^P=1^-$ in ${\bf\bar 3}$ representation are
\begin{align}\label{mass}
 m(T^{cc}_{ud}({\bf \bar 3}))&= 3.82 {\rm GeV} , & J^P=1^- ,
 \\
  m(T^{cc}_{us}({\bf \bar 3}))=  m(T^{cc}_{ds}({\bf \bar 3}))&=4.07 {\rm GeV} , & J^P=1^- .
\end{align}
The masses of the tetraquarks with  the spin-parity $J^P=1^-$ in ${\bf 6}$ representation are
\begin{align}
 m(T^{cc}_{ud}({\bf 6}))&= \left\{ \begin{array} {ll}
  4.21 {\rm GeV} , & J^P=1^- , \\
4.19 {\rm GeV} , & J^P=1^- ,
 \\ 4.14 {\rm GeV}, & J^P=1^- ,\end{array} \right.\\
  m(T^{cc}_{us}({\bf 6}))&= \left\{ \begin{array} {ll}
 4.39 {\rm GeV} , & J^P=1^- , \\
4.36{\rm GeV} , & J^P=1^- ,
 \\ 4.31 {\rm GeV}, & J^P=1^- ,\end{array} \right.\\
  m(T^{cc}_{ss}({\bf 6}))&= \left\{ \begin{array} {ll}
  4.58 {\rm GeV} , & J^P=1^- , \\
4.56 {\rm GeV} , & J^P=1^- ,
 \\ 4.51{\rm GeV}, & J^P=1^- .\end{array} \right.\label{mass3}
\end{align}

\section{Doubly charmed tetraquarks decays to  two charmed mesons\label{IV}}

When the doubly charmed tetraquarks lie above the ${D\bar D}$ threshold, they can decay into two charmed mesons.  For the tetraquarks with positive parity, the two-body decay amplitudes can be written as:
\begin{align}
{\cal M}(T^{cc}[0^+] \to \bar{D} \bar{D})&=  F_{\bar{D}\bar{D}}f_{T^{cc}},
\end{align}
where $f_{T^{cc}}$ is the decay constant of the tetraquark,
 $F_{\bar{D} \bar{D}}$ denotes the effective coupling to diquark-antidiquark pair.

For the tetraquarks with $J^P=1^{--}$, the two body  decay amplitudes are written as:
\begin{align}
{\cal M}(T^{cc}[1^{--}] \to \bar D \bar{D})= & \varepsilon_{T^{cc}} \cdot (P_{\bar{D}} -P'_{\bar{D}}) \frac{F_{\bar D\bar{D}}f'_{T^{cc}}}{3^{1/2}m_{T_{c\bar{c}}} },\\
{\cal M}(T^{cc}[1^{--}] \to \bar D \bar{D}^*)=& \varepsilon^{\mu}_{T_{c\bar{c}}} \varepsilon^{\nu}_{\bar{D}^*} P^\rho_{\bar{D}} P'^\sigma_{\bar{D}} \epsilon_{\mu\nu\rho\sigma}\nonumber\\
&F_{\bar D\bar{D}^*}\frac{f'_{T^{cc}}}{3^{1/2}(m_{T_{c\bar{c}}})^{2} },
\end{align}
\begin{align}
&{\cal M}(T^{cc}[1^{--}] \to \bar D^*(P) \bar{D}^*(P'))\nonumber\\
= & \varepsilon^{\mu}_{T_{c\bar{c}}}\varepsilon^\rho_{\bar{D}^*}\varepsilon'^\nu_{\bar D^*}  \frac{F_{\bar D^*\bar{D}^*}f_{T^{cc}}}{3^{1/2}m_{T^{cc}} }(g^{\mu\rho}(-P_T-P_{\bar{D}^*})^\nu\nonumber\\
&+g^{\mu\nu}P_T^\rho+g^{\mu\nu}{P'}_{\bar D^*}^\rho+g^{\rho\nu}({P'}_{\bar{D}^*}-P_{\bar D^*})^\mu),
\end{align}

The decay width of $T^{cc} \to \bar D( \bar D^*)\bar{D}(\bar{D}^*)$ can be written as:
\begin{eqnarray}
\Gamma(T^{cc} \to \bar D( \bar D^*)+\bar{D}(\bar{D}^*))&=&\frac{|\textbf{p}|}{8\pi
m_{T^{cc}}^2}|{\cal M}|^2,
\end{eqnarray}
where
\begin{eqnarray}
|\textbf{p}|&=&\frac{\sqrt{\left(m_{T^{cc}}^2-\left(m_1-m_{2}\right){}^2\right) \left(m_{T^{cc}}^2-\left(m_{1}+m_{2}\right){}^2\right)}}{2 m_{T^{cc}}},\nonumber
\end{eqnarray}
which is the momentum modulus of final charmed meson in the tetraquark rest frame. $m_1$ and $m_2$
are the final charmed meson's masses, respectively.
\begin{figure}[th]
\begin{center}
\includegraphics[width=0.38\textwidth]{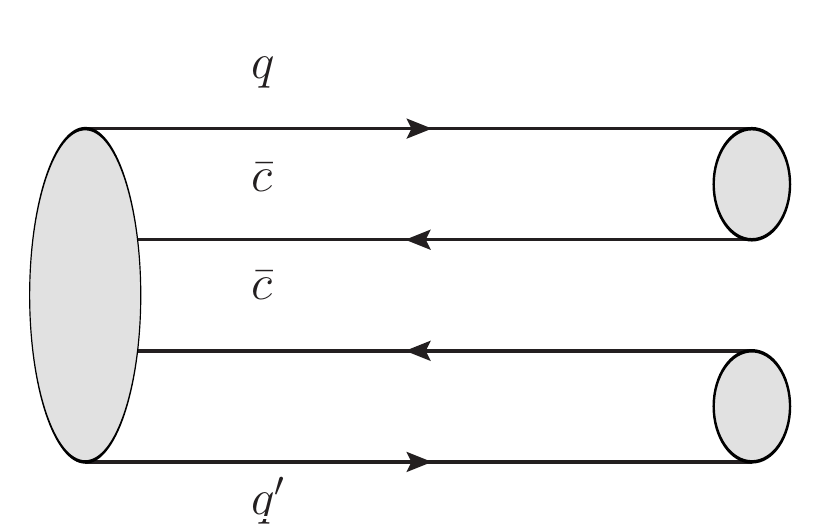}
\end{center}
    \vskip -0.7cm \caption{Feynman Diagrams for doubly charmed tetraquarks decays to  two charmed mesons. }\label{Fig:f0p-x}
\end{figure}

The decay ratio for the tetraquark with positive parity is
\begin{eqnarray}
\frac{\Gamma(T^{cc}[0^+] \to \bar D \bar{D})}{F_{\bar D\bar{D}}^2(f_{T^{cc}})^2|\textbf{p}|}&=&\frac{1}{8\pi
m_{T^{cc}}^2}.\nonumber
\end{eqnarray}

The similar ratios for the tetraquarks with $J^P=1^{--}$ are
\begin{eqnarray}
\frac{\Gamma(T^{cc}[1^{--}] \to \bar D \bar{D})}{F_{\bar D\bar{D}}^2(f'_{T^{cc}})^2|\textbf{p}|^3}&=&\frac{1}{6\pi
m_{T^{cc}}^4},\nonumber\\
\frac{\Gamma(T^{cc}[1^{--}] \to\bar D \bar{D}^*)}{F_{ \bar D \bar{D}^*}^2(f'_{T^{cc}})^2|\textbf{p}|^3}&=&\frac{1}{12\pi
m_{T^{cc}}^4},\nonumber\\
\frac{\Gamma(T^{cc}[1^{--}] \to\bar D^* \bar{D}^*)}{F_{ \bar D^* \bar{D}^*}^2(f'_{T^{cc}})^2|\textbf{p}|^3}&=&\frac{m_{T^{cc}}^4-
\frac{104}{9}m_{T^{cc}}^2|\textbf{p}|^2+\frac{48}{9}|\textbf{p}|^4}{2\pi
m_{T^{cc}}^4(m_{T^{cc}}^2-4|\textbf{p}|^2)^2}.\nonumber\\
\end{eqnarray}

The decay widths of $1^{--}$ tetraquarks to double charmed mesons will be suppressed compared to
that of the tetraquarks with positive parity to double charmed mesons.  The suppression factor for the  $\bar D\bar{D}$ final states can be estimated as
\begin{eqnarray}
\frac{\Gamma(T^{cc}[1^{--}] \to \bar D \bar{D})}{\Gamma(T^{cc}[0^+] \to \bar D \bar{D})}&\simeq &\frac{4(f'_{T^{cc}})^2|\textbf{p}|^2}{3(f_{T^{cc}})^2 m_{T^{cc}}^2},
\end{eqnarray}

The decay constant of the  doubly charmed tetraquark $f'_{T^{cc}}$ are not easily to extract clearly through the hadronic decays. These nonperturbative information can be extract through the purely leptonic decays as $T^{cc}\to 2\ell^-+ 2\bar{\nu_\ell}$ for the doubly charged tetraquarks, $T^{cc}\to 2\ell^- +\ell^++\bar{\nu_\ell}$ for the singly charged tetraquarks, and $T^{cc}\to 2\ell^- +2\ell^+$ for the neutral tetraquarks.

 \section{Doubly charmed tetraquarks decays to  a charmed baryon and a  light baryon\label{IV}}

In this section, we will study the doubly charmed tetraquarks weak decays into  baryon and anti-baryon, where one of baryons is charmed.
Due to larger phase space, these decays may provide more useful information to discover the doubly charmed tetraquarks at experiment.

The charm quark decays into light quarks are classified into three types: $c\to s\bar{d} u$,
$c\to u\bar{d} d/\bar{s} s$, and $c\to d\bar{s} u$, namely Cabibbo-favored, singly Cabibbo-suppressed, and
doubly Cabibbo-suppressed topologies, respectively. The Cabibbo-favored channels may provide a windows to discover the doubly charmed tetraquarks.

At the hadron level, the effective Hamiltonian in charm decays behave as the certain representations
under the flavor SU(3) symmetry as ${\bf\bar 3}\otimes {\bf3}\otimes {\bf\bar
3}={\bf\bar 3}\oplus {\bf\bar 3}\oplus {\bf6}\oplus {\bf\overline{15}}$. So the Hamiltonian  can
be decomposed in terms of a vector $H^i({\bf\overline3})$, a traceless
tensor antisymmetric in upper indices, $H^{[ij]}_k({\bf6})$, and a
traceless tensor symmetric in   upper indices,
$H^{\{ij\}}_k({\bf\overline{15}})$.

For the Cabibbo-favored $c\to s\bar{d} u$ decay, we have the nonzero components~\cite{Wang:2017azm}
\begin{eqnarray}
 H^{31}_2({\bf6})&=&-H^{13}_2({\bf6})=1,\nonumber\\
 H^{31}_2({\bf \overline{15}})&=& H^{13}_2({\bf \overline{15}})=1,
\end{eqnarray}
where the representation $H^i({\bf \bar 3})$ will vanish in the
SU(3) flavor symmetry.

Singly charmed baryons with two light quarks can form an anti-triplet or sextet, which can be classified into two matrixes
\begin{eqnarray}
 T^{c}_{\bf{\bar 3}}=  \frac{1}{\sqrt{2}} \left(\begin{array}{ccc} 0 & \Lambda_c^+  &  \Xi_c^+  \\ -\Lambda_c^+ & 0 & \Xi_c^0 \\ -\Xi_c^+   &  -\Xi_c^0  & 0
  \end{array} \right),
\end{eqnarray}
\begin{eqnarray}
 T^c_{\bf{6}} = \left(\begin{array}{ccc} \Sigma_c^{++} &  \frac{1}{\sqrt{2}}\Sigma_c^+   & \frac{1}{\sqrt{2}} \Xi_c^{\prime+}\\
  \frac{1}{\sqrt{2}}\Sigma_c^+& \Sigma_c^{0} & \frac{1}{\sqrt{2}} \Xi_c^{\prime0} \\
  \frac{1}{\sqrt{2}} \Xi_c^{\prime+}   &  \frac{1}{\sqrt{2}} \Xi_c^{\prime0}  & \Omega_c^0
  \end{array} \right)\,.
\end{eqnarray}

The light baryons made of three light quarks  with  spin-parity $J^P=\frac{1}{2}^+$ can form  an  octet, which has the expression
\begin{eqnarray}
T_8= \left(\begin{array}{ccc} \frac{1}{\sqrt{2}}\Sigma^0+\frac{1}{\sqrt{6}}\Lambda^0 & \Sigma^+  &  p  \\ \Sigma^-  &  -\frac{1}{\sqrt{2}}\Sigma^0+\frac{1}{\sqrt{6}}\Lambda^0 & n \\ \Xi^-   & \Xi^0  & -\sqrt{\frac{2}{3}}\Lambda^0
  \end{array} \right).
\end{eqnarray}

\begin{figure}[th]
\begin{center}
\includegraphics[width=0.36\textwidth]{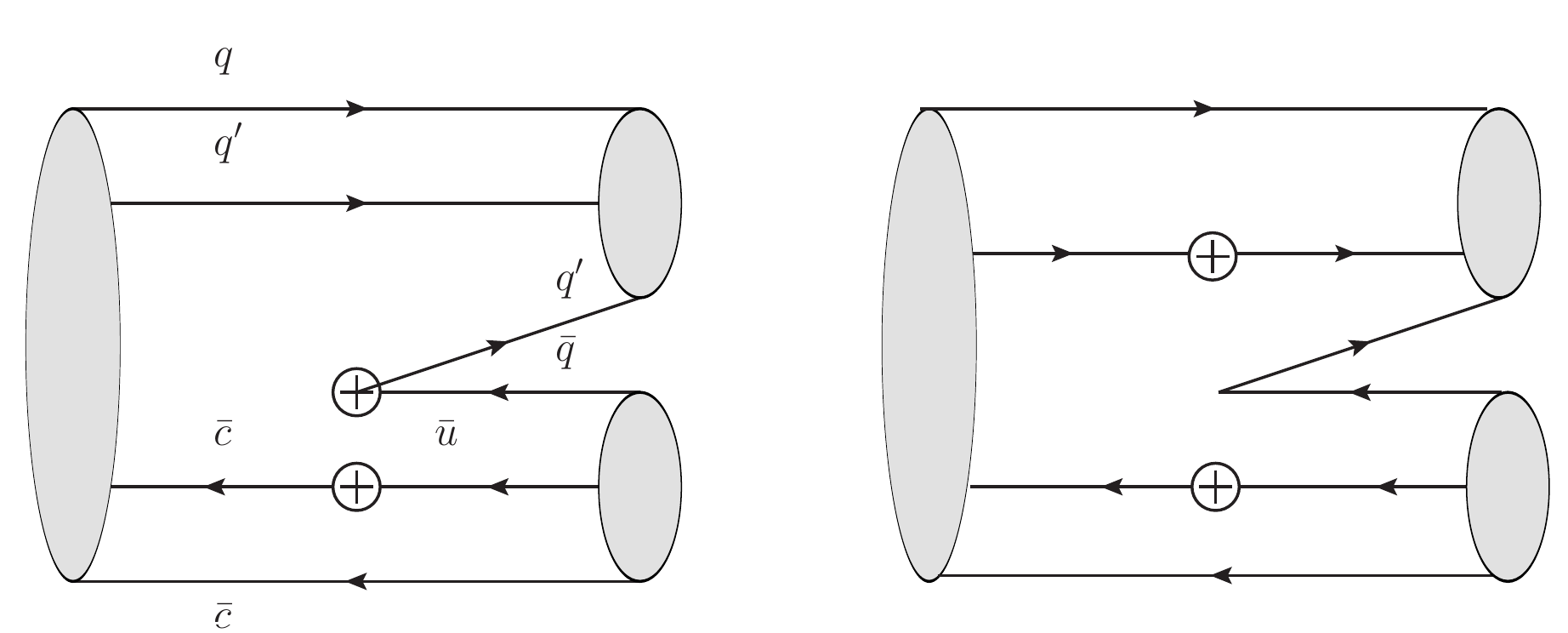}
\end{center}
    \vskip -0.7cm \caption{Feynman Diagrams for doubly charmed tetraquarks decays to  baryon and anti-baryon. The  two circles ``$\oplus$~$\oplus$'' denote the four fermion weak effective vertex.}\label{Fig:f0p-x}
\end{figure}

The decay amplitudes of Cabibbo-favored channels $A(T^{cc}\to  \bar  T^{c}+ T_8) = \langle \bar  T^{c}+ T_8\vert {\cal H}_{eff} \vert T^{cc} \rangle$ can be written as $V_{cs}V^*_{ud} A^T(T^{cc}\to  \bar  T^{c}+ T_8)$.

For the decays $T^{cc}({\bf \bar{3}} )\to   \bar T^{c}_{\bf{\bar 3}}+ T_8$,
the effective Hamiltonian can be written as
\begin{eqnarray}
 &&A^T(T^{cc}( {\bf \bar{3}} )\to  \bar T^{c}_{\bf{\bar 3}}+  T_8)\nonumber\\
 &=& a_6 T^{cc}_{[ij]}({\bf \bar{3} }) { T^{c}_{\bf{\bar 3}}}_{[k,l]} \epsilon^{ikn} (\bar T_8)^m_n H^{[jl]}_m ( {\bf6})\nonumber\\
 &&+ a'_6 T^{cc}_{[ij]}({\bf \bar{3} }) { T^{c}_{\bf{\bar 3}}}_{[k,l]} \epsilon^{imn} (\bar T_8)^k_n H^{[jl]}_m ( {\bf6})\nonumber\\
 &&+ b_6 T^{cc}_{[ij]}({\bf \bar{3} }) { T^{c}_{\bf{\bar 3}}}_{[k,l]} \epsilon^{imn} (\bar T_8)^j_n H^{[kl]}_m ( {\bf6})\nonumber\\
 &&+a_{15} T^{cc}_{[ij]}({\bf \bar{3} }) { T^{c}_{\bf{\bar 3}}}_{[k,l]} \epsilon^{ikn} (\bar T_8)^m_n H^{\{jl\}}_m ( {\bf \overline{ 15}})\nonumber\\
 &&+ a'_{15} T^{cc}_{[ij]}({\bf \bar{3} }) { T^{c}_{\bf{\bar 3}}}_{[k,l]} \epsilon^{imn} (\bar T_8)^k_n H^{\{jl\}}_m ( {\bf\overline{ 15}}).\label{eq:amp1}
\end{eqnarray}

Similarly, the effective Hamiltonian have the same formulae for the decays $T^{cc}({\bf 6} )\to  \bar T^{c}_{\bf{\bar 3}}+  T_8$ by the replacement of $T^{cc}_{[ij]}({\bf \bar{3} }) \to T^{cc}_{\{ij\}}({\bf 6} )$. The term proportional to $ b_6$ will vanish for both the $T^{cc}({\bf \bar{3} })\to  \bar T^{c}_{\bf{6}}+  T_8$ and $T^{cc}({\bf 6} )\to  \bar T^{c}_{\bf{6}}+  T_8$  decays.
The Cabibbo-allowed amplitudes for doubly charmed tetraquarks decays to  a charmed baryon and a  light baryon are given in Tabs.~\ref{amp33}, \ref{amp36}, \ref{amp63} and \ref{amp66}. For convenience, we give the  ratio of the decay widths of the doubly charmed tetraquarks to a charmed anti-baryon and a light baryon in the flavor SU(3) symmetry in Tabs.~\ref{decayWidth3} and \ref{decayWidth6}.

\begin{table}
\caption{Decay amplitudes of $T^{cc}({\bf \bar{3}} )\to  \bar T^{c}_{\bf{\bar 3}}+  T_8$ decays.  }\begin{tabular}{|c|c|c|c|}\hline\hline
Decay channel & amplitude  \\\hline

$ T^{cc}_{ud}({\bf \bar{3}}) \to \bar \Xi_c^- n $ & $\frac{1}{2}(a_{15}-a_6-2b_{6})$
\\\hline

$ T^{cc}_{ds} ({\bf \bar{3}}) \to \bar \Xi_c^- \Sigma^-$ & $-\frac{1}{2}(a_{15}+a_6+2b_{6})$
\\\hline
$ T^{cc}_{us} ({\bf \bar{3}}) \to \bar\Xi_c^- \Sigma^0$ & $\frac{1}{2\sqrt{2}}(a'_{15}-2a_6-a'_6-2b_{6})$
\\\hline
$ T^{cc}_{us} ({\bf \bar{3}}) \to \bar \Xi_c^- \Lambda^0$ & $\frac{1}{2\sqrt{6}}(3a'_{15}+2a_6+a'_6+2b_{6})$
\\\hline
$ T^{cc}_{us} ({\bf \bar{3}}) \to \bar \Xi_c^0 \Sigma^-$ & $\frac{1}{2}(a_{15}+a'_{15}-a_6-a'_6)$
\\\hline
$ T^{cc}_{us} ({\bf \bar{3}}) \to \bar \Lambda_c^-  n$ & $-\frac{1}{2}(a_{15}+a'_{15}+a_6+a'_6)$
\\
\hline\hline
\end{tabular} \label{amp33}
\end{table}

\begin{table}
\caption{Decay amplitudes of $T^{cc}({\bf \bar{3}} )\to  \bar T^{c}_{\bf{ 6}}+ T_8$ decays.  }\begin{tabular}{|c|c|c|c|}\hline\hline
Decay channel & amplitude  \\\hline

$ T^{cc}_{ud}({\bf \bar{3}}) \to \bar\Sigma_c^{'-}  n $ & $\frac{1}{2}(a_{15}-a_6)$
\\\hline
$ T^{cc}_{ud}({\bf \bar{3}}) \to \bar \Omega_c^0  \Sigma^- $ & $-\frac{1}{\sqrt{2}}(a_{15}-a_6)$
\\\hline
$ T^{cc}_{ds} ({\bf \bar{3}}) \to \bar \Sigma_c^{--}  n $ &  $-\frac{1}{\sqrt{2}}(a_{15}+a_6)$
\\\hline
$ T^{cc}_{ds} ({\bf \bar{3}}) \to  \bar \Sigma_c^{'-}\Sigma^- $ &  $\frac{1}{2}(a_{15}+a_6)$
\\\hline
$ T^{cc}_{us} ({\bf \bar{3}}) \to \bar \Sigma_c^-  n$ & $\frac{1}{2}(a_{15}+a'_{15}+a_6+a'_6)$
\\\hline
$ T^{cc}_{us} ({\bf \bar{3}}) \to \bar \Sigma_c^{--}  p+ $ & $\frac{1}{\sqrt{2}}(a'_{15}+a'_6)$
\\\hline
$ T^{cc}_{us} ({\bf \bar{3}}) \to \bar \Sigma_c^{'-}  \Lambda^0 $ & $-\frac{1}{2\sqrt{6}}(2a_{15}+a'_{15}+3a'_6)$
\\\hline
$ T^{cc}_{us} ({\bf \bar{3}}) \to \bar \Sigma_c^{'-}  \Sigma^0 $ & $\frac{1}{2\sqrt{2}}(2a_{15}+a'_{15}-a'_6)$
\\\hline
$ T^{cc}_{us} ({\bf \bar{3}}) \to \bar \Omega_c^{0}  \Xi^- $
& $\frac{1}{\sqrt{2}}(a'_{15}-a'_6)$
\\\hline
$ T^{cc}_{us} ({\bf \bar{3}}) \to \bar \Sigma_c^{'0} \Sigma^- $
& $\frac{1}{2}(a_{15}+a'_{15}-a_6-a'_6)$
\\
\hline\hline
\end{tabular} \label{amp36}
\end{table}
\begin{table}
\caption{Decay amplitudes of $T^{cc}({\bf 6} )\to  \bar T^{c}_{\bf{\bar 3}}+  T_8$ decays.  }\begin{tabular}{|c|c|c|c|}\hline\hline
Decay channel & amplitude  \\\hline

$ T^{cc}_{ud}({\bf 6}) \to \bar \Xi_c^- n $ & $-\frac{1}{2}(a_{15}-a_6+2b_{6})$
\\\hline
$ T^{cc}_{ds} ({\bf 6}) \to \bar \Xi_c^- \Sigma^-$ & $-\frac{1}{2}(a_{15}+a_6-2b_{6})$
\\\hline
$ T^{cc}_{us} ({\bf 6}) \to \bar\Xi_c^- \Sigma^0$ & $\frac{1}{2\sqrt{2}}(-2a_{15}-a'_{15}+a'_6+2b_{6})$
\\\hline
$ T^{cc}_{us} ({\bf 6}) \to \bar \Xi_c^- \Lambda^0$ & $\frac{1}{2\sqrt{6}}(2a_{15}+a'_{15}+3a'_6+6b_{6})$
\\\hline
$ T^{cc}_{us} ({\bf 6}) \to \bar \Xi_c^0 \Sigma^-$ & $-\frac{1}{2}(a_{15}+a'_{15}-a_6-a'_6)$
\\\hline
$ T^{cc}_{us} ({\bf 6}) \to \bar \Lambda_c^-  n$ & $-\frac{1}{2}(a_{15}+a'_{15}+a_6+a'_6)$
\\\hline
$ T^{cc}_{uu} ({\bf 6}) \to \bar \Xi_c^0  n$ & $\frac{1}{\sqrt{2}}(a_{15}+a'_{15}-a_6-a'_6)$
\\\hline
$ T^{cc}_{uu} ({\bf 6}) \to \bar \Xi_c^-  p$ & $\frac{1}{\sqrt{2}}(a'_{15}-a'_6-2b_6)$
\\\hline
$ T^{cc}_{ss} ({\bf 6}) \to \bar \Xi_c^-  \Xi^- $ & $\frac{1}{\sqrt{2}}(a'_{15}+a'_6+2b_6)$
\\\hline
$ T^{cc}_{ss} ({\bf 6}) \to \bar \Lambda_c^-  \Sigma^- $ & $\frac{1}{\sqrt{2}}(a_{15}+a'_{15}+a'_6+b_6)$
\\\hline
\hline\hline
\end{tabular} \label{amp63}
\end{table}

\begin{table}
\caption{Decay amplitudes of $T^{cc}({\bf 6} )\to  \bar T^{c}_{\bf{ 6}}+ T_8$ decays.  }\begin{tabular}{|c|c|c|c|}\hline\hline
Decay channel & amplitude  \\\hline

$ T^{cc}_{ud}({\bf 6}) \to \bar\Sigma_c^{'-}  n $ & $-\frac{1}{2}(a_{15}-a_6)$
\\\hline
$ T^{cc}_{ud}({\bf6}) \to \bar \Omega_c^0  \Sigma^- $ & $\frac{1}{\sqrt{2}}(a_{15}-a_6)$
\\\hline
$ T^{cc}_{ds} ({\bf 6}) \to \bar \Sigma_c^{--}  n $ &  $-\frac{1}{\sqrt{2}}(a_{15}+a_6)$
\\\hline
$ T^{cc}_{ds} ({\bf 6}) \to  \bar \Sigma_c^{'-}\Sigma^- $ &  $\frac{1}{2}(a_{15}+a_6)$
\\\hline
$ T^{cc}_{us} ({\bf 6}) \to \bar \Sigma_c^-  n$ & $\frac{1}{2}(a_{15}+a'_{15}+a_6+a'_6)$
\\\hline
$ T^{cc}_{us} ({\bf 6}) \to \bar \Sigma_c^{--}  p+ $ & $\frac{1}{\sqrt{2}}(a'_{15}+a'_6)$
\\\hline
$ T^{cc}_{us} ({\bf 6}) \to \bar \Sigma_c^{'-}  \Lambda^0 $ & $-\frac{1}{2\sqrt{6}}(3a'_{15}+2a_6+a'_6)$
\\\hline
$ T^{cc}_{us} ({\bf 6}) \to \bar \Sigma_c^{'-}  \Sigma^0 $ & $-\frac{1}{2\sqrt{2}}(a'_{15}-2a_6-a'_6)$
\\\hline
$ T^{cc}_{us} ({\bf 6}) \to \bar \Omega_c^{0}  \Xi^- $
& $-\frac{1}{\sqrt{2}}(a'_{15}-a'_6)$
\\\hline
$ T^{cc}_{us} ({\bf 6}) \to \bar \Sigma_c^{'0} \Sigma^- $
& $-\frac{1}{2}(a_{15}+a'_{15}-a_6-a'_6)$
\\\hline
$ T^{cc}_{uu} ({\bf 6}) \to \bar \Sigma_c^{'0} n $
& $\frac{1}{\sqrt{2}}(a_{15}+a'_{15}-a_6-a'_6)$
\\\hline
$ T^{cc}_{uu} ({\bf 6}) \to \bar \Sigma_c^{'-} p^+ $
& $\frac{1}{\sqrt{2}}(a'_{15}-a'_6)$
\\\hline
$ T^{cc}_{uu} ({\bf 6}) \to \bar \Omega_c^{0} \Lambda^0 $
& $\frac{1}{\sqrt{6}}(-a_{15}-2a'_{15}+a_6+2a'_6)$
\\\hline
$ T^{cc}_{uu} ({\bf 6}) \to \bar \Omega_c^{0}\Sigma^0 $
& $\frac{1}{\sqrt{2}}(a_{15}-a_6)$
\\\hline
$ T^{cc}_{ss} ({\bf 6}) \to \bar \Sigma_c^{--} \Lambda^0 $
& $\frac{1}{\sqrt{6}}(a_{15}-a'_{15}+a_6-a'_6)$
\\\hline
$ T^{cc}_{ss} ({\bf 6}) \to \bar \Sigma_c^{--} \Sigma^0 $
& $-\frac{1}{\sqrt{2}}(a_{15}+a'_{15}+a_6+a'_6)$
\\\hline
$ T^{cc}_{ss} ({\bf 6}) \to \bar \Sigma_c^{'-} \Xi^- $
& $-\frac{1}{\sqrt{2}}(a'_{15}+a'_6)$
\\\hline
$ T^{cc}_{ss} ({\bf 6}) \to \bar \Sigma_c^{-} \Sigma^- $
& $-\frac{1}{\sqrt{2}}(a_{15}+a'_{15}+a_6+a'_6)$
\\
\hline\hline
\end{tabular} \label{amp66}
\end{table}

\begin{table}
\caption{SU(3) relations for decay widths for the ${\bf \bar{3}}$ doubly charmed tetraquark to a charmed anti-baryon and a light baryon.   $R$ denotes
the ratio of two decay widths.}\begin{tabular}{|c|c|c|c|}\hline\hline
$ \Gamma(channel_1)/\Gamma(channel_2) $ & $R$  \\\hline
$ \frac{\Gamma( T^{cc}_{ud}({\bf \bar{3}}) \to \bar \Omega_c^0  \Sigma^- )}{\Gamma(T^{cc}_{ud}({\bf \bar{3}}) \to \bar\Sigma_c^{'-}  n)} $ & 2
\\\hline
$ \frac{\Gamma( T^{cc}_{ds} ({\bf \bar{3}}) \to \bar \Sigma_c^{--}  n )}{\Gamma(T^{cc}_{ds} ({\bf \bar{3}}) \to  \bar \Sigma_c^{'-}\Sigma^-) } $ & 2
\\\hline
$ \frac{\Gamma( T^{cc}_{us} ({\bf \bar{3}}) \to \bar \Sigma_c^{--}  p+  )}{\Gamma(T^{cc}_{ds} ({\bf \bar{3}}) \to  \bar \Sigma_c^{'-}\Sigma^-) } $ & 2
\\\hline
$ \frac{\Gamma( T^{cc}_{us} ({\bf \bar{3}}) \to \bar \Sigma_c^-  n)}{\Gamma(T^{cc}_{us} ({\bf \bar{3}}) \to \bar \Lambda_c^-  n) } $ & 1
\\\hline
$ \frac{\Gamma( T^{cc}_{us} ({\bf \bar{3}}) \to \bar \Xi_c^0 \Sigma^-)}{\Gamma(T^{cc}_{us} ({\bf \bar{3}}) \to \bar \Sigma_c^{'0} \Sigma^-) } $ & 1
\\
\hline\hline
\end{tabular} \label{decayWidth3}
\end{table}

\begin{table}
\caption{SU(3) relations for decay widths for the ${\bf 6}$  doubly charmed tetraquark to a charmed anti-baryon and a light baryon.   $R$ denotes
the ratio of two decay widths.}\begin{tabular}{|c|c|c|c|}\hline\hline
$ \Gamma(channel_1)/\Gamma(channel_2) $ & $R$  \\\hline
$ \frac{\Gamma( T^{cc}_{ss} ({\bf 6}) \to \bar \Lambda_c^-  \Sigma^-)}{\Gamma(T^{cc}_{us} ({\bf 6}) \to \bar \Lambda_c^-  n)} $ & 2
\\\hline
$  \frac{\Gamma( T^{cc}_{us} ({\bf 6}) \to \bar \Sigma_c^-  n)}{\Gamma(T^{cc}_{us} ({\bf 6}) \to \bar \Lambda_c^-  n)}  $ & 1
\\\hline
$  \frac{\Gamma( T^{cc}_{ss} ({\bf 6}) \to \bar \Sigma_c^{--} \Sigma^0)}{\Gamma(T^{cc}_{us} ({\bf 6}) \to \bar \Lambda_c^-  n)}  $ & 2
\\\hline
$  \frac{\Gamma( T^{cc}_{ss} ({\bf 6}) \to \bar \Sigma_c^{-} \Sigma^-)}{\Gamma(T^{cc}_{us} ({\bf 6}) \to \bar \Lambda_c^-  n)}  $ & 2
\\\hline
$ \frac{\Gamma( T^{cc}_{uu} ({\bf 6}) \to \bar \Xi_c^0  n)}{\Gamma(T^{cc}_{us} ({\bf 6}) \to \bar \Xi_c^0 \Sigma^-) } $ & 2
\\\hline
$ \frac{\Gamma( T^{cc}_{us} ({\bf 6}) \to \bar \Sigma_c^{'0} \Sigma^-)}{\Gamma(T^{cc}_{us} ({\bf 6}) \to \bar \Xi_c^0 \Sigma^-) } $ & 1
\\\hline
$ \frac{\Gamma( T^{cc}_{uu} ({\bf 6}) \to \bar \Sigma_c^{'0} n)}{\Gamma(T^{cc}_{us} ({\bf 6}) \to \bar \Xi_c^0 \Sigma^-) } $ & 2
\\\hline
$ \frac{\Gamma(T^{cc}_{ud}({\bf6}) \to \bar \Omega_c^0  \Sigma^- )}{\Gamma(T^{cc}_{ud}({\bf 6}) \to \bar\Sigma_c^{'-}  n) } $ & 2
\\\hline
$ \frac{\Gamma(T^{cc}_{us} ({\bf 6}) \to \bar \Omega_c^{0}  \Xi^- )}{\Gamma(T^{cc}_{ud}({\bf 6}) \to \bar\Sigma_c^{'-}  n) } $ & 2
\\\hline
$ \frac{\Gamma(T^{cc}_{uu} ({\bf 6}) \to \bar \Sigma_c^{'-} p^+ )}{\Gamma(T^{cc}_{ud}({\bf 6}) \to \bar\Sigma_c^{'-}  n) } $ & 2
\\\hline
$ \frac{\Gamma(T^{cc}_{uu} ({\bf 6}) \to \bar \Omega_c^{0}\Sigma^0 )}{\Gamma(T^{cc}_{ud}({\bf 6}) \to \bar\Sigma_c^{'-}  n) } $ & 2
\\\hline
$ \frac{\Gamma(T^{cc}_{ds} ({\bf 6}) \to \bar \Sigma_c^{--}  n )}{\Gamma(T^{cc}_{ds} ({\bf 6}) \to  \bar \Sigma_c^{'-}\Sigma^-) } $ & 2
\\\hline
$ \frac{\Gamma(T^{cc}_{us} ({\bf 6}) \to \bar \Sigma_c^{--}  p+)}{\Gamma(T^{cc}_{ds} ({\bf 6}) \to  \bar \Sigma_c^{'-}\Sigma^-) } $ & 2
\\\hline
$ \frac{\Gamma(T^{cc}_{ss} ({\bf 6}) \to \bar \Sigma_c^{'-} \Xi^- )}{\Gamma(T^{cc}_{ds} ({\bf 6}) \to  \bar \Sigma_c^{'-}\Sigma^-) } $ & 2
\\
\hline\hline
\end{tabular} \label{decayWidth6}
\end{table}

Whatever the doubly charmed tetraquarks decay to two charmed mesons or decay to a charmed baryon and a light baryon, BESIII, BelleII, and LHCb  are excellent experiment platforms to search for it. BESIII has accumulated large $e^+e^-$ collision data samples at an energy range from 3.8 GeV to 4.6 GeV and will continue taking data at open-charm energy region~\cite{Ablikim:datataking}. It would be an interesting research at those energy points with the world's top integrated luminosity just as papers~\cite{Ablikim:opencharm1,Ablikim:opencharm2,Ablikim:opencharm3,Ablikim:opencharm4} et al. The two charmed mesons channels $T^{cc} \to \bar D( \bar D^*)+\bar{D}(\bar{D}^*)$ can be studied through the whole energy region from 3.8 GeV to 4.6 GeV. A charmed baryon and a light baryon decay channels showed in Tabs.~\ref{amp33}, \ref{amp36}, \ref{amp63} and \ref{amp66} can also be studied carefully at the same energy region.

\section{Conclusion}
 We have considered the possibility for the doubly charmed tetraquark states with the quark configuration $[qq'][\bar c\bar c]$. These states are  straightforward consequences of the constituent diquark-antidiquark model. The doubly charmed tetraquarks form the antitriplet and sextet configuration in the flavor SU(3) symmetry. The mass spectrum and their spin-parities of tetraquark states have been investigated. We found that a doubly charmed tetraquark $T^{cc}_{ud}({\bf \bar 3})$ with spin-parity $J^P=1^+$ is around 140MeV below the ${D\bar D}$ threshold and 270MeV below the ${D\bar D^*(D^*\bar D)}$ threshold. For $T^{cc}[0^+]$ and $T^{cc}[1^{--}]$ tetraquark states, the decay modes of the two-body charmed mesons have also been presented. Furthermore, the doubly charmed tetraquarks decays to  a charmed baryon and a  light baryon  have been studied in the SU(3) flavor symmetry.  The search for such kind of states can be carried out at BESIII, BelleII, LHCb and other experiments with their large data samples of high luminosity. These Cabibbo-favored channels shall provide a widows to discover the possible doubly charmed tetraquark states.

\section*{Acknowledgments}
This work was supported in part  by Natural Science Foundation of
Jiangsu under Grant No.~BK20171471, by the National Natural Science Foundation of China under Grant No.~U1732105, and by the Research Foundation for
Advanced Talents of Nanjing Normal University under Grant No.~2014102XGQ0085.

\end{document}